\documentclass[a4paper,12pt, epsfig]{article}
\usepackage{epsfig}
\usepackage{amssymb}
\usepackage{amsfonts}
\usepackage{amsmath}
\usepackage{graphicx}

\newskip\humongous \humongous=0pt plus 1000pt minus 1000pt

\newif\ifdtup

\catcode`\@=11

\@addtoreset{equation}{section}
\def\theequation{\thesection.\arabic{equation}}

\def\@normalsize{\@setsize\normalsize{15pt}\xiipt\@xiipt
\abovedisplayskip 14pt plus3pt minus3pt%
\belowdisplayskip \abovedisplayskip
\abovedisplayshortskip \z@ plus3pt%
\belowdisplayshortskip 7pt plus3.5pt minus0pt}

\def\small{\@setsize\small{13.6pt}\xipt\@xipt
\abovedisplayskip 13pt plus3pt minus3pt%
\belowdisplayskip \abovedisplayskip
\abovedisplayshortskip \z@ plus3pt%
\belowdisplayshortskip 7pt plus3.5pt minus0pt
\def\@listi{\parsep 4.5pt plus 2pt minus 1pt
     \itemsep \parsep
     \topsep 9pt plus 3pt minus 3pt}}

\relax

\catcode`@=12

\catcode`\@=11

\def\section{\@startsection{section}{1}{\z@}{3.5ex plus 1ex minus
   .2ex}{2.3ex plus .2ex}{\large\bf}}

\def\thesection{\arabic{section}}
\def\thesubsection{\arabic{section}.\arabic{subsection}}

\def\appendix{\setcounter{section}{0}
 \def\thesection{Appendix \Alph{section}}
 \def\thesubsection{\Alph{section}.\arabic{subsection}}
 \def\theequation{\Alph{section}.\arabic{equation}}}

\def\SymBoxes#1#2#3#4{\newdimen\un@t \un@t#3%
\raisebox{#1}{\rule{#2\un@t}{#4}\hskip-#2\un@t
\@tempdimb\un@t \advance\@tempdimb by-#4\@tempcntb#2\relax%
\@whilenum{\@tempcntb>0}\do{
\rule{#4}{\un@t}\hskip\@tempdimb \advance\@tempcntb by\m@ne}%
\hskip-#2\un@t \rule[\un@t]{#2\un@t}{#4}%
\rule[\un@t]{#4}{#4}\hskip-#4
\rule{#4}{\un@t}}\hskip-#4}                

\begin{document}

\newcommand{\beq}{\begin{equation}}
\newcommand{\eeq}{\end{equation}}
\newcommand{\bea}{\begin{eqnarray}}
\newcommand{\eea}{\end{eqnarray}}
\newcommand{\beas}{\begin{eqnarray*}}
\newcommand{\eeas}{\end{eqnarray*}}
\newcommand{\defi}{\stackrel{\rm def}{=}}
\newcommand{\non}{\nonumber}
\newcommand{\bquo}{\begin{quote}}
\newcommand{\enqu}{\end{quote}}
\renewcommand{\(}{\begin{equation}}
\renewcommand{\)}{\end{equation}}
\def \eqn#1#2{\begin{equation}#2\label{#1}\end{equation}}
\def\IZ{{\mathbb Z}}
\def\IR{{\mathbb R}}
\def\IC{{\mathbb C}}
\def\IQ{{\mathbb Q}}
\def\de{\partial}
\def\Tr{ \hbox{\rm Tr}}
\def\H{ \hbox{\rm H}}
\def\HE{ \hbox{$\rm H^{even}$}}
\def\HO{ \hbox{$\rm H^{odd}$}}
\def\K{ \hbox{\rm K}}
\def\Im{ \hbox{\rm Im}}
\def\Ker{ \hbox{\rm Ker}}
\def\const{\hbox {\rm const.}}
\def\o{\over}
\def\im{\hbox{\rm Im}}
\def\re{\hbox{\rm Re}}
\def\bra{\langle}\def\ket{\rangle}
\def\Arg{\hbox {\rm Arg}}
\def\Re{\hbox {\rm Re}}
\def\Im{\hbox {\rm Im}}
\def\exo{\hbox {\rm exp}}
\def\diag{\hbox{\rm diag}}
\def\longvert{{\rule[-2mm]{0.1mm}{7mm}}\,}
\def\a{\alpha}
\def\dag{{}^{\dagger}}
\def\tq{{\widetilde q}}
\def\p{{}^{\prime}}
\def\W{W}
\def\N{{\cal N}}
\def\hsp{,\hspace{.7cm}}

\def\br{\nonumber\\}
\def\IZ{{\mathbb Z}}
\def\IR{{\mathbb R}}
\def\IC{{\mathbb C}}
\def\IQ{{\mathbb Q}}
\def\IP{{\mathbb P}}
\def \eqn#1#2{\begin{equation}#2\label{#1}\end{equation}}

\newcommand{\C}{\ensuremath{\mathbb C}}
\newcommand{\Z}{\ensuremath{\mathbb Z}}
\newcommand{\R}{\ensuremath{\mathbb R}}
\newcommand{\rp}{\ensuremath{\mathbb {RP}}}
\newcommand{\cp}{\ensuremath{\mathbb {CP}}}
\newcommand{\vac}{\ensuremath{|0\rangle}}
\newcommand{\vact}{\ensuremath{|00\rangle}                    }
\newcommand{\oc}{\ensuremath{\overline{c}}}
\begin{titlepage}
\begin{flushright}
SISSA 28/2010/EP
\end{flushright}
\bigskip
\def\thefootnote{\fnsymbol{footnote}}

\begin{center}
{\Large
{\bf
Black Hole Vacua and Rotation\\
\vspace{0.1in}
}
}
\end{center}

\bigskip
\begin{center}
{\large  Chethan
KRISHNAN}\\
\end{center}

\renewcommand{\thefootnote}{\arabic{footnote}}

\begin{center}
{\em  { SISSA,\\
Via Bonomea 265, 34136, Trieste, Italy\\
{\rm {\texttt{krishnan@sissa.it}}}\\}}

\end{center}

\noindent
\begin{center} {\bf Abstract} \end{center}
Recent developments suggest that the near-region of rotating black holes behaves like a CFT. To understand this better, I propose to study quantum fields in this region. An instructive approach for this might be to put a large black hole in AdS and to think of the entire geometry as a toy model for the ``near-region". Quantum field theory on rotating black holes in AdS can be well-defined (unlike in flat space), if fields are quantized in the co-rotating-with-the-horizon frame. First, some generalities of constructing Hartle-Hawking Green functions in this approach are discussed. 
Then as a specific example where the details are easy to handle, I turn to 2+1 dimensions (BTZ), write down the Green functions explicitly starting with the co-rotating frame, and observe some structural similarities they have with the Kerr-CFT scattering amplitudes. Finally, in BTZ, there is also an alternate construction for the Green functions: we can start from the covering $AdS_3$ space and use the method of images. Using a 19th century integral formula, I show the equality between the boundary correlators arising via the two constructions. 

\begin{center}
\end{center}

\vspace{1.6 cm}
\vfill

\end{titlepage}
\hfill{}
\bigskip

\tableofcontents

\setcounter{footnote}{0}
\section{Preliminaries: Motivation and Setup}

\noindent

Classical black hole mechanics is formally identical to the laws of thermodynamics \cite{BCH}. One way to make this analogy physical is to consider an eternal\footnote{The only kind of black hole we will study.} black hole \cite{HH1}. In flat space, an eternal black hole  can be described by its Hartle-Hawking vacuum, which is the thermal state capturing the fact that the evaporation from the black hole has come to equilibrium with a heat bath at the Hawking temperature. This would be the natural way to treat a thermal state in the canonical ensemble (which the static black hole is believed to be). But since black holes in flat space have negative specific heat, the potential equilibrium between the black hole and the thermal bath is necessarily an unstable one.

The situation gets worse for spinning black holes. Again here, black hole mechanics has a thermodynamical description, but in the grand canonical ensemble where the angular velocity acts as the chemical potential \cite{GPP}. But unlike in the static case, there is no good construction for a Hartle-Hawking vacuum \cite{FT, Kay}. This has to do with the fact that there is no Killing vector field that is timelike everywhere outside the horizon, and so there is no good way of quantizing fields with a definite notion of positive energy. A related, but distinct problem is that the spacetime suffers from super-radiant instabilities.

In anti-de Sitter (AdS) space black holes are much better defined than in flat space. In particular, large  black holes have positive specific heat and can be stable. A large Euclidean AdS black hole has lesser action than thermal AdS and therefore dominates the partition function \cite{Hawking-Page, Witten}. This thermal nature of the black hole is also visible in Lorentzian signature. One can view a large static eternal black hole in AdS as a specific entangled (thermal) state between two CFTs, where the CFTs live on the two asymptotic boundaries of the maximally extended geometry \cite{Israel, Maldacena}. Indeed, the fact that black holes have a good thermal description in AdS is really a foregone conclusion these days: many of the recent developments in ``applied" AdS/CFT are nothing but dual versions of black hole physics.

This raises the  question: is AdS the preferred context to understand spinning black holes as well? The purpose of this paper is to explore this question.

Our primary motivation in investigating this question comes from recent work on the Kerr-CFT correspondence \cite{CMS}. The idea here is that the ``near-region" of a generic (i.e., possibly far-from-extremal) Kerr black hole  has a hidden conformal symmetry. It was found that the scattering amplitudes of fields on the black hole in this limit have a structure that can be interpreted as arising from a CFT at finite left and right temperatures. The entropy of the black hole can also be reproduced by related arguments. In this picture, the black hole near-region  responds to external probes (scalar fields) like a CFT would, at the appropriate temperatures. Since Kerr black holes are expected to be physical, it is worthwhile investigating this intriguing picture in detail.

As a natural first step, we would like to try to quantize fields in this background. In the original AdS/CFT correspondence,  the map between bulk and boundary correlators was a very useful tool in understanding the physics, and we would like to see what kind of information can be extracted in the case of Kerr-CFT by studying Green functions in the near-region. In the Kerr-CFT case there is an auxiliary $AdS_3$ space (unrelated to the physical geometry at least at first blush) that seems to be important in understanding the near-region of the black hole. The black hole geometry breaks this $AdS_3$ symmetry due to periodic identifications on the azimuthal circle. It is not clear what the full relevance of this AdS is. Along the same lines lies the problem that even though there seems to be a CFT, it is not clear if it can be associated with a clean ``boundary". In particular, unlike in the usual (static) extremal black holes, here there is no simple AdS throat.

When one tries to quantize fields on asymptotically flat Kerr black holes, one runs into the immediate problem mentioned above concerning the absence of satisfactory timelike coordinates. The state of the art regarding the quantization of scalars in Kerr spacetime is captured in \cite{FT, Winstanley, Ottewill}. A manifestation of the absence of a good time coordinate everywhere outside the horizon is that a frame co-rotating with the horizon becomes superluminal far away from the black hole. One of the conclusions of \cite{Ottewill} is that if one truncates the geometry by putting the black hole inside a mirror that lies within this velocity-of-light surface, then one can construct Hartle-Hawking Green functions and stress tensors (at least numerically). It is crucial for this approach that the mirror be close enough to the black hole \cite{Ottewill}.

One of the motivations of this paper is that a black hole in a small enough box might be an instructive model for the near-region of the Kerr black hole. To get an idea, note that the condition that the box is not too large can be written as
\bea
M \sim r_0
\eea
where $M$ is the mass of the black hole and $r_0$ is the radius to the boundary of the box, which can be taken to be at the velocity-of-light surface. In particular, this statement is independent of the frequency of the mode under consideration. On the contrary, the near-region is defined for each mode in the geometry as the regime where the relations
\bea
M \ll \frac{1}{\omega}, \ \  r \ll \frac{1}{\omega},
\eea
hold. From the form of these relations, it is clear that as far as the low energy dynamics is considered, the physics of the two cases have a regime of overlap. So optmistically, one might be able to understand aspects of one system by studying the other.

In turn, a very natural way to describe some of the relevant physics of a small enough box around a black hole is to consider a large black hole in AdS with reflecting boundary conditions. 
For a large black hole in AdS, 
\bea
M\sim r_0 \equiv L
\eea
where $L$ is the AdS scale. Unlike in flat space, AdS avoids a velocity-of-light surface in the co-rotating frame not by a cut-off at a finite radial coordinate, but because of the AdS warp factor. The radial coordinate can run to infinity, but particles can bounce back from the boundary in finite time. Despite the enormous amount of work done on AdS black holes in the last years, it seems that a systematic effort towards defining quantum field theory on Kerr black holes in AdS has not been made. Some of the material that I present here is possibly part of the lore, but since most of it is not explicitly stated in publicized work, it is perhaps worthwhile to collect it in one place and save someone else the trouble of re-inventing the wheel.

Generically, AdS has a stabilizing effect on large black holes. But the issue of super-radiance in the case of rotating black holes is more subtle. Superradiance is an instability that happens because certain modes get repeatedly amplified by absorbing energy from the hole because they are trapped in its gravitational well. In flat space, massless particles are not trapped, but in AdS they can be reflected from the boundary in finite time, so even they are trapped. This is again roughly a consequence of the fact that AdS is like a reflecting box. So one might suspect that super-radiant instabilities are going to be worse for AdS black holes. Fortunately, this is not the case. There are simple general arguments by Hawking and Reall \cite{HawkReall} that show that there is no super-radiant instability for matter that satisfies the dominant energy condition when the black hole is large and rotating slowly enough. The question of super-radiant instabilities was addressed in terms of the relevant length scales in \cite{Cardoso} and arguments were made that large Kerr-AdS black holes are stable \footnote{Small AdS black holes are well-established to be unstable to super-radiance \cite{SR1}.}. A full fledged gauge-invariant perturbation theory analysis has been done in \cite{Murata} and indeed stable regimes have been found. But this was done for doubly spinning $AdS_5 \times S^5$ black holes, while our concern is primarily singly spinning black holes in this paper \footnote{But it seems quite possible that the discussion here might have generalizations to the doubly spinning case as well. In \cite{CK2} it was shown that very general, in particular multi-spinning, black holes are likely to have a good Kerr-CFT description. See \cite{rest} for related work.}. In \cite{Kodama2} such an analysis was made for singly rotating black holes, but unfortunately (due to technical reasons) the analysis was limited to dimensions greater than 6.  But the encouraging result of the analysis was indeed that stable black holes exist in the regimes predicted by Hawking and Reall \cite{HawkReall}. The review \cite{lefteris} contains some discussion on regimes of stability of various kinds of black holes in various dimensions. A nice discussion of superradiance in the context of AdS black holes can be found in \cite{Kodama2}. In any event, it is implicitly assumed that the black holes we consider in this paper are stable. The general consensus seems to be that at least for slowly rotating large AdS black holes, both perturbative and thermodynamical stability are expected. We expect that as long as the black hole is classically stable, we can do quantization around the background, so the arguments of this paper should hold. If the black  hole is not classically (perturbatively) stable, then I suspect that the methods here should be discarded. Incidentally, it is worth keeping in mind that even though the flat space Kerr black hole is unstable to super-radiance, there is still a useful quasi-stationary notion of a spinning black hole because the time-scale of the instability is very large. The Green functions that we write down might have some use in this kind of  classically unstable but quasi-stationary cases as well, but further investigation is needed for a definitive answer.

For rotating AdS black holes, quantizing fields in the co-rotating frame is a natural thing to do because this time coordinate is timelike and Killing everywhere outside the horizon. We describe some basic aspects of the mode constructions that arise and contrast that to the work of Frolov and Thorne \cite{FT} in the flat space case. In particular the wave equation is separable as in flat space. There are some technicalities which make the explicit computations complicated for Kerr-$AdS_4$ (closely related to some {\em technical} problems that also exist in ordinary Kerr), so we only present the general aspects here\footnote{A full treatment will require a (numerical?) solution of the wave equation in Kerr-AdS, analogous to the numerical work done in \cite{Ottewill} for flat space Kerr.}. Using these, we describe how to construct Green functions in the Hartle-Hawking-like vacuum. We find the natural emergence of the grand canonical ensemble with angular velocity as the chemical potential. There is an associated entangled CFT interpretation \cite{Maldacena}.

Two of the big complications for Kerr-$AdS_4$ is that (1) the wave equation is not solvable in a useful way, and (2) the metric has a polar angle (roughly\footnote{``Roughly", because the black hole is rotating and the harmonics are no longer spherical, in fact in AdS, they are not even spheroidal.} the $\theta$ on $S^2$) dependence which might not be separable in an asymptotically AdS coordinate system. 
We can bypass both these difficulties while retaining the essence of the problem if we work with spinning black holes  in $AdS_3$, a la BTZ. Here the radial scalar equation is solvable in terms of hypergeometric functions and the only angle is the azimuthal angle, so there are no complications. The thermal (Hartle-Hawking) Green functions can be explicitly written down by quantizing in the co-rotating frame and using standard techniques of quantum field theory in curved spacetime. After writing them down, we briefly comment on the structural similarities that these Green functions have with the recently discussed scattering amplitudes in Kerr-CFT.  Another observation is that  in 2+1 dimensions,  there is an alternate construction for the black hole Green functions from the covering $AdS_3$ space. After an interesting Fourier transformation, we find that the two Green functions give rise to the same boundary correlators. This will be taken as further evidence that the construction is ``natural". We comment also on the possibility that the near-region Kerr Green functions might be obtained via an appropriate quotienting of the auxiliary $AdS_3$.

{\bf Other Questions:} Our immediate motivation in this paper was the recent work on Kerr-CFT \cite{CMS} and  the near-region physics of black holes. But to emphasize the generality of some of these questions, we list some open problems.

First and foremost, a full-fledged study of the Green functions and stress tensor renormalization on rotating AdS black holes is clearly of interest in the context of AdS/CFT. Some of this work will necessarily have to be numerical \cite{Ottewill}, but general arguments could also lead to some insight \cite{Winstanley}.

A possible alternative application of the construction here might be in understanding the internal structure of black holes applying the methods of \cite{Kraus, Shenker, Liu, CK1}. The idea is that the analytic continuation of the correlators can probe the interior of the geometry. Correlators, in the large mass limit (large compared to the AdS scale, not the Planck mass), are dominated by bulk geodesics. Therefore the task is to identify the  map between correlators and geodesics so that one can identify the correlators that probe the relevant parts of the geometry. For regions inside the horizon, these are certain spacelike geodesics. There are systematic techniques \cite{Liu} for identifying them. With the Hartle-Hawking correlator at hand, this is in principle straightforward.
Unfortunately the wave equation is generically not solvable in higher (than three) dimensional Kerr-AdS geometries.

In \cite{CK1} the problem was addressed in the case of the BTZ  black hole (where the wave equation is solvable) by exploiting a coordinate system that simplified the problem. In higher dimensions, things cannot be expected to be so simple. But in the large-mass limit, the solvability of the wave equation might not be deadly because what is important are the pole structure of the Green functions, and there exits strategies for (approximately) determining these quasi-normal poles \cite{Liu2}.

Another possible way to bypass some of the complications is to work with black branes instead of black holes in AdS. In this case, we still have non-trivial local curvature, but the horizon has the topology $\IR \times \IR^{d-2} \times S^1$ instead of the $\IR \times S^{d-1}$ of the Kerr-$AdS_{d+1}$ black hole. The first $\IR$ in both cases stands for the time direction. Metrics for spinning black branes have been written down in \cite{Lemos} and they turn out to be related in a simple way to the metrics of non-rotating black holes. It should be straightforward to translate the quasinormal pole computations of static black holes to this case. It is also interesting that the internal structure of black branes is different from that of black holes: in particular, the analogue of the Cauchy horizon is a radius (a turning point) beyond which the physical radius increases. Studying the internal structure of these black branes using boundary correlators is a project that is currently under way.

Yet another possible direction of development is the generalization of the work here to the many black objects that have recently been constructed in higher dimensions \cite{citedump2}. Recent work on Kerr-CFT is collected in \cite{citedump1}.

\section{$AdS_{4}$}


In this section we will work with a Kerr-$AdS_{d+1}$ black hole with $d=3$. The statements we make here should have straightforward generalizations in higher dimensions as well, at the very least when the black hole is singly spinning. The radial scalar equation in this geometry is not solvable in a useful form analytically, and we will never use explicit expressions for the mode solutions\footnote{Exact solutions are only available in $d=2$, the BTZ case, which we will use in the next section.}. The explicit metric for Kerr-AdS in Boyer-Lindquist (which is a Schwarzschild-like coordinate system) is \cite{Carter2}
\bea
ds^{2} =
  -\frac {\Delta _{r}}{\rho ^{2}} \left[
  dt - \frac {a \sin ^{2} \theta }{\Sigma } d\phi \right] ^{2}
  + \frac {\rho ^{2}}{\Delta _{r}} dr ^{2}
  + \frac {\rho ^{2}}{\Delta _{\theta }} d\theta ^{2}
  + \frac {\Delta _{\theta }\sin ^{2} \theta }{\rho ^{2}} \left[
  a \, dt - \frac {r^{2}+a^{2}}{\Sigma } d\phi \right] ^{2},
\label{eq:metric}
\eea
with
\bea
\rho ^{2} = r^{2}+a^{2} \cos ^{2} \theta , \ \  \Sigma =
1- \frac {a^{2}}{l^{2}} , \hspace{0.8in}
\\
  \Delta _{r} =
 \left( r^{2}+ a^{2} \right) \left( 1 + \frac {r^{2}}{l^{2}} \right) - 2Mr , \ \
  \Delta _{\theta } =
1 -\frac{a^{2}}{l^{2}} \cos ^{2} \theta .
\eea
Here $a$ is an angular momentum parameter, $M$ is a mass parameter and $l$ is the AdS radius.
A computation reveals that the wave equation separates in these coordinates like it did for flat space Kerr \cite{Carter, Winstanley2}:
\bea
\label{static}
u_{\omega,p}(t,\phi,x;r)=
\frac{1}{\sqrt{2\omega}}e^{-i\omega t+i p \phi}Y_l(\theta)X_{\omega,p,l}(r)
\eea
We work with singly rotating black holes with rotation direction $\phi$. The polar angle $\theta$ of the $S^{2}$ is captured in terms of the harmonics $Y_l(\theta)$, which are more complicated than even the spheroidal form found for flat space Kerr. This seems to suggest that the separation constants can only be determined numerically, throwing a spanner on any hope for analytic determination of Green functions. $X_{\omega,p,l}$ are the solutions of the radial part of the scalar field equation in Kerr-AdS. The $\frac{1}{\sqrt{2\omega}}$ is a convention \cite{Birrell}: we want the Green functions to have a more-or-less standard form when we assume reflective boundary conditions at the timelike boundary so that the Cauchy problem is well-defined, and demand that the Klein-Gordon inner product on a spacelike slice is normalized to unity \cite{Birrell}. This does not fix the normalization uniquely, because one can absorb normalizations into the definition of $X_{\omega, p,l}$. Our choice\footnote{I thank G. Festuccia for correspondence on this.} corresponds to setting $X_{\omega,p,l}$ to be of the form $e^{\pm i \omega z}$ at the horizon where $z$ is the tortoise coordinate (see \cite{Liu} and section 4.2 of \cite{CK1}). Explicit radial and angular equations can be found in \cite{Winstanley2}.

The trouble, as already mentioned is that the standard Boyer-Lindquist ``time" is not well-defined everywhere outside the horizon. But unlike in flat space, (rotating) black holes in AdS have a Killing vector that is everywhere timelike outside the horizon \cite{HawkReall, CK1}. This time is the time coordinate in a frame that is co-rotating with the horizon:
\bea
\frac{\partial }{\partial T}={\partial \over \partial t}+\Omega {\partial \over \partial \phi}.
\eea
where $\Omega=\frac{a \Sigma}{r_+^2 +a^2}$ is the horizon angular velocity as seen in Boyer-Lindquist. Here $r_+$ is the outer horizon determined as the bigger root of $\Delta_r=0$. As long as the angular velocity parameter\footnote{\label{omega}There is a slight subtlety due to the fact that in the Boyer-Lindquist form that we have written down, the angular velocity of the horizon naively seems to be $\Omega=\frac{a \Sigma}{r_+^2 +a^2}$ and not $\Omega_0$ above. This is misleading because BL is an asymptotically rotating frame. In a coordinate system that is asymptotically AdS and non-rotating, the horizon has an angular velocity given by $\Omega_0$. See \cite{Marc}, and the discussion about Henneaux-Teitelboim coordinates later in this section.} of the metric $\Omega_0$ satisfies $\Omega_0<1$, with
\bea
\Omega_0\equiv\frac{a (1+r_+^2/l^2)}{r_+^2 +a^2},
\eea
the spacetime is stable against super-radiance if the matter propagating in the geometry satisfies some reasonable energy conditions \cite{HawkReall} (See also the discussion in the previous section on super-radiance.)
Because of the AdS warp factor, the rigid co-rotation does {\em not} become superluminal anywhere. So we will quantize scalars in this frame and a positive norm basis for the Hilbert space can be obtained by restricting the Klein-Gordon inner product to positive frequency modes. A simple coordinate choice where this Killing vector is the time coordinate is
\bea
T=t, \ \ \Phi=\phi-\Omega t, \ \ 
\eea
Note that at fixed $T$, $\Phi$ has the same periodicity as $\phi$ had at fixed $t$, namely $2\pi$\footnote{To belabor this elementary point, the identification is $(t,\phi) \sim (t,\phi+2\pi)$ in the original coordinates. This means that $(t, \phi-\Omega t) \sim (t, \phi-\Omega t+2\pi)$ for any $\Omega$.}. This means that $p$ is quantized in integers as before.
The modes become
\bea
u_{\omega_-,p,l}(T,\Phi,\theta;r)=
\frac{1}{\sqrt{2\omega_-}}e^{-i\omega_- T+i p \Phi}Y_l(\theta)X_{\omega_-,p,l}(r). \label{co-modes}
\eea
The connection with the previous modes is that
\bea
e^{i p \phi}e^{-i\omega t}=e^{ip\Phi}e^{-i(\omega- p \Omega)T}\equiv e^{ip\Phi}e^{-i\omega_- T}.
\eea
This mode-expansion makes sense for quantization, as opposed to the static one in (\ref{static}). The positive energy modes are those with $\omega_- > 0$. In terms of this, the scalar field operator can be expanded in the usual way as \cite{Birrell}
\bea
X(T,\Phi,\theta;r)=\sum_{l,\omega_-,p}(a_{l,\omega_-,p}u_{l,\omega_-,p}+a^\dagger_{l,\omega_-,p}u^*_{l,\omega_-,p}).
\eea
with $\omega_-$ restricted to positive values alone. The creation and annihilation operators $a, a^\dagger$ satisfy the usual algebra.

\begin{figure}
\begin{center}
\includegraphics[height=0.4\textheight
]{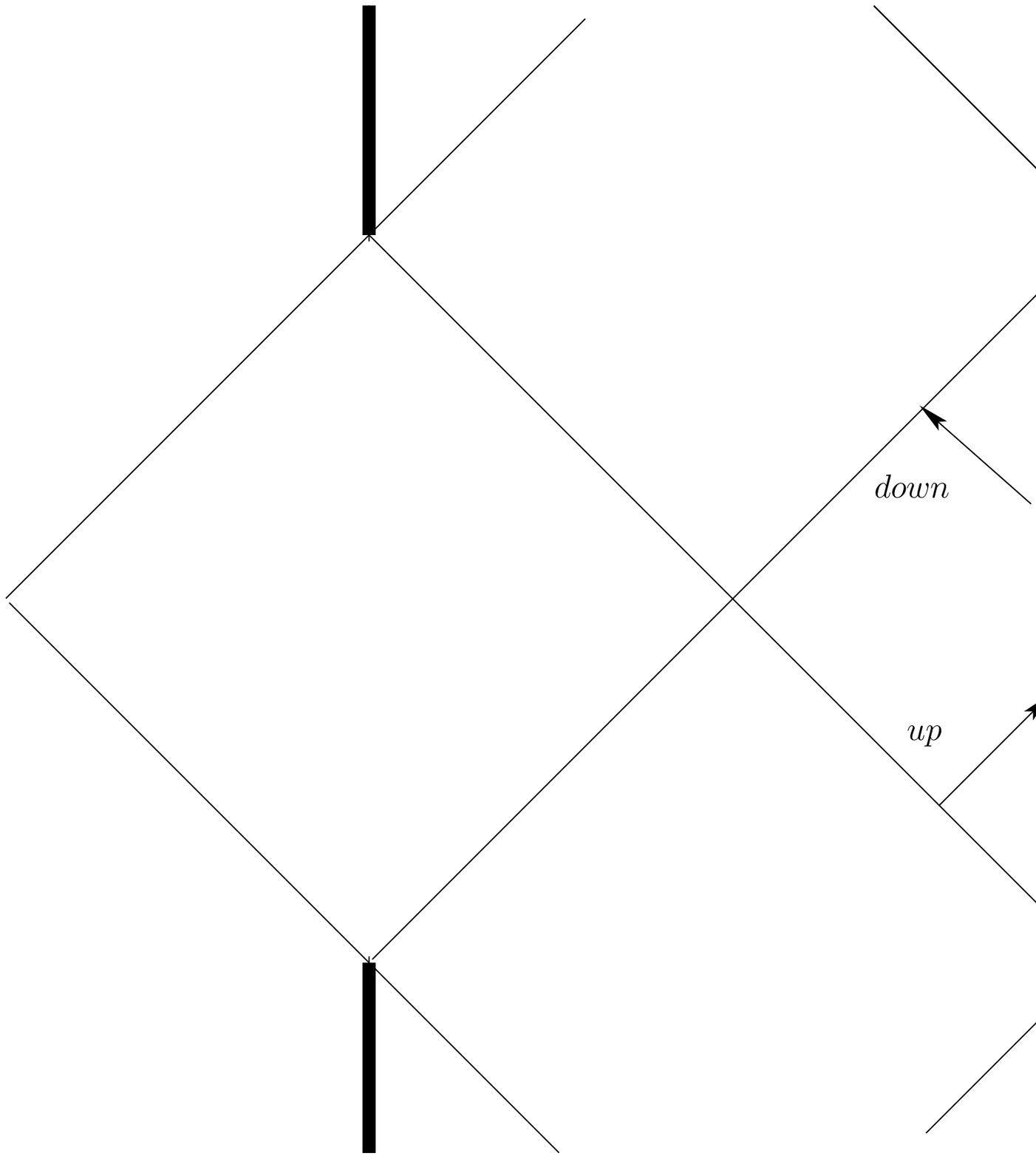}
\caption{Schematic Penrose diagram of a Kerr black hole in flat space.}
\label{penflat}
\end{center}
\end{figure}

\begin{figure}
\begin{center}
\includegraphics[height=0.4\textheight
]{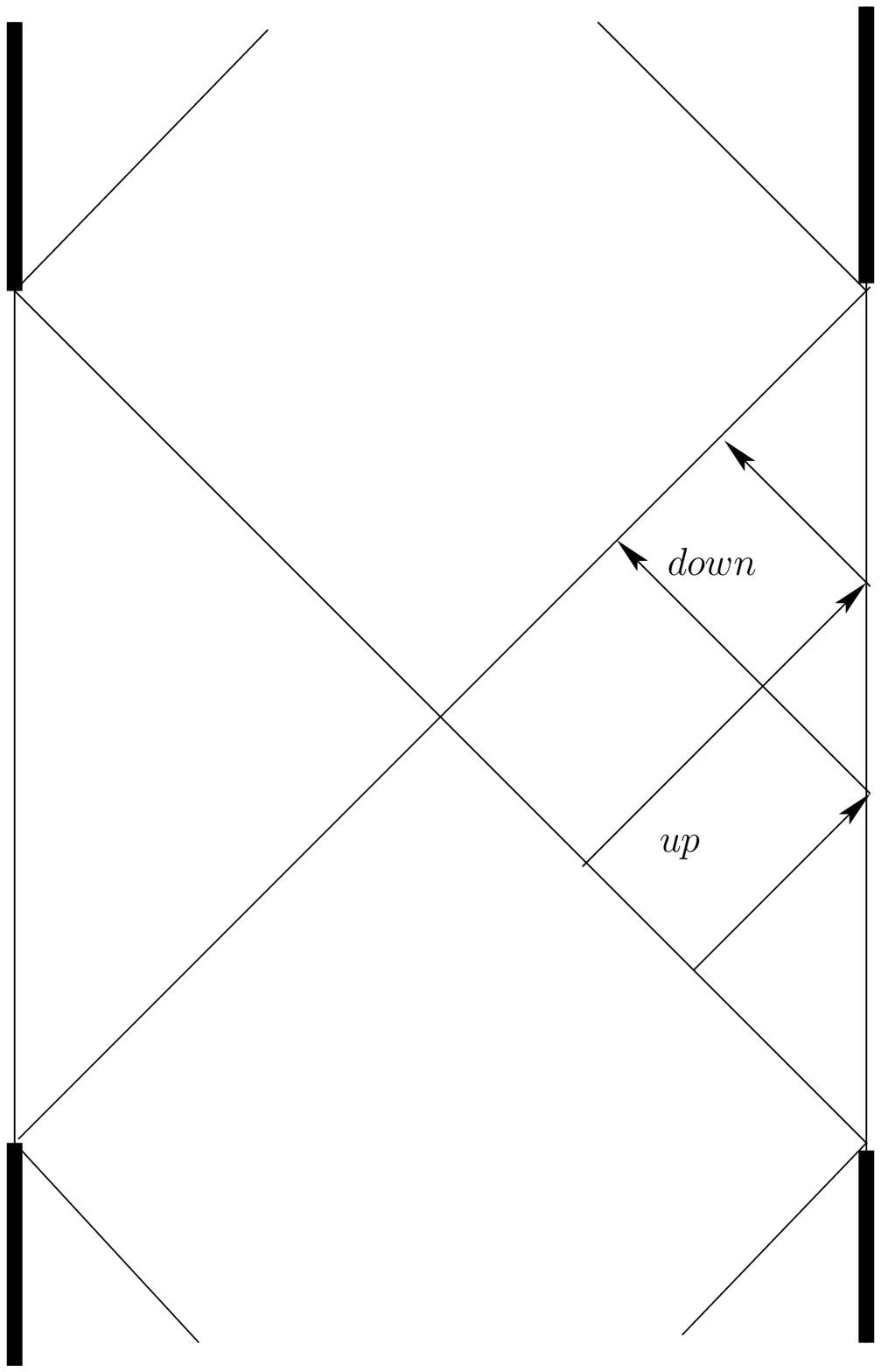}
\caption{Schematic Penrose diagram of a Kerr-AdS black hole}
\label{penk}
\end{center}
\end{figure}

Now we turn to some comments about the nature of the modes. First, note that we can introduce tortoise coordinates (see eg. \cite{Liu, CK1, Winstanley, Winstanley2}) to write the Kerr radial equation in a Schr\"odinger form with a potential. On general grounds (as well as the explicit expression in \cite{Winstanley2}), we know that the potential in these coordinates has an infinite barrier at the AdS boundary for a scalar with big enough\footnote{Note that scalar masses can be negative in AdS. The massless case is already big enough for our purposes, see for eg., \cite{Winstanley2}.} mass. The solution is unique when we demand normalizability. Near the horizon, the effect of the cosmological constant is negligible, and the potential barrier  is finite as in flat space. Note that the overall structure of the potential is therefore very different from that in flat space Kerr, where the potential went to a finite constant at both the horizon and at infinity. In flat space, the presence of superradiant modes makes it difficult to define a consistent notion of positive energy. This lead to the introduction in \cite{FT} of various kinds of modes and ``viewpoints". In the co-rotating frame in AdS on the other hand, with the natural AdS fall-offs at the boundary, it is possible to see that the modes at the horizon (i.e., the $X_{\omega_-,p}$) are a combination of ingoing and outgoing waves (basically because the potential is finite there). These are the analogues of the ``up" and ``down" modes in the ``near-horizon viewpoint", in the language of \cite{FT}.  Figures \ref{penflat}, \ref{penk} show the distinction between the modes introduced in flat space and AdS using the relevant parts of the Penrose diagrams of the black holes.
In other words, normalizable boundary conditions give rise to outgoing waves at the past horizon and ingoing waves at the future horizon. 
In flat space, typically one can also allow the possibility that there are modes coming in from past null infinity (``in" modes) and going out into future null infinity (``out" modes). The existence of a general mode expansion with appropriate positivity of energy that can provide a basis for all of these physically distinct phenomena is what gives rise to the complications (and possible inconsistency) in flat space Kerr black hole. In the AdS case on the other hand, the natural boundary conditions are those that allow reflections at the boundary (here is where the similarity with the reflecting box comes in). Physically the picture is that an evaporating large black hole can come to equilibrium with its own Hawking radiation, when the boundary is reflecting.

Note that in \cite{Avis} a different set of boundary conditions  (``transparent") were also considered for fields in AdS, but they are less natural from the conventional AdS/CFT point of view. Transparent boundary conditions lead to leakage into the CFT, which we would like to avoid. Note also that when we allow leakage, we are giving up the advantages of the mirror at the boundary and we are back to worrying about super-radiance and related issues familiar from the flat space Kerr case. 

The modes we found are solutions in a Schwarzschild-like coordinate system. The Hartle-Hawking-like vacuum that we are after is defined in terms of modes in Kruskal coordinates. So we can use the general strategy for constructing Green functions on bifurcate Killing horizons using a Bogolubov transformation analogous to the one used for demonstrating the Unruh effect in Rindler space \cite{Birrell, Kay}. 
We can write down the Bogolubov transformations from the knowledge that the translation to Kruskal coordinates is implemented via
\bea
U_+=\pm e^{-\kappa_+u}, \ \ V_+=e^{\kappa_+ v}.
\eea
The notation here is standard, see e.g., \cite{Poisson}. The precise coordinates needed in this construction above for the Kerr-AdS case are a simple generalization of those written down in a remarkable paper by Carter 42 years ago for the flat case \cite{Carter}. In terms of the Hawking temperature $T_H\equiv \frac{1}{\beta_H}=\frac{2\pi}{\kappa_+}$. The Bogolubov transformations only depend on the surface gravity $\kappa_+$ of the outer horizon, and so we will not present the details of the coordinate change\footnote{They are obtained by a trivial modification of the expressions that lead to Eqn. (26) in section 1.C of \cite{Carter}.}. In practice, the construction of the Green function amounts to thermally populating the up modes described above at the Hawking temperature.


In flat space Kerr with standard boundary conditions, it is known that there cannot be a good HH vacuum\footnote{The precise statement is that there cannot be a stationary Hadamard state that is regular everywhere and is invariant under simultaneous $(t,\phi)$ reversal.} because of a theorem due to Kay and Wald \cite{Kay}. But when one truncates the space with a mirror before the the speed of light surface, the stress tensor and the Hartle-Hawking vacuum become well-defined \cite{Winstanley, Ottewill}. AdS with reflecting boundary conditions is a natural mirror of this kind, and so it stands to reason that there is a well-defined Hartle-Hawking vacuum.

In defining a mirror, one needs to define its shape; in particular one has to describe its profile in terms of the angles $\theta$ and $\phi$. The numerical construction in flat space done in \cite{Ottewill} did this by declaring that the mirror was at a fixed $r=r_0$ with no angular dependence. This issue is a bit subtle in our case. This is because we would like to set asymptotically AdS boundary conditions as the definition of our mirror, while our Boyer-Lindquist coordinate system is (as emphasized in footnote \ref{omega}) not quite asymptotically of the static AdS form. In other words, we need to set our AdS boundary conditions in terms of the correct radial coordinate. A simple solution to this problem would be to use the Henneaux-Teitelboim coordinates $(\tau,\varphi,y, \Theta)$ \cite{Marc} where the asymptotically AdS nature of the metric is manifest:
\bea
\tau=t, \qquad  \varphi = \phi + \frac {a}{l^{2}} t, \qquad  y \cos \Theta = r \cos \theta, \qquad  y^{2} = \frac {1}{\Sigma } \left[ r^{2} \Delta _{\theta } + a^{2} \sin ^{2} \theta \right] \label{HT}.
\eea
Here $y$ is an asymptotically AdS radial variable. Unfortunately, the metric becomes rather complicated in this coordinate system and it is not clear to me whether the wave equation can be separated as in Boyer-Lindquist. Conceptually though, this picture offers a clean solution with standard AdS boundary conditions: if one has a (numerical) solution for the $y, \Theta$ part of the equation of motion, we can go to the co-rotating frame as before via
\bea
T=\tau, \ \ \Phi=\varphi-\Omega_0 \tau,
\eea
and the rest of the arguments are unaffected. The boundary will be at $y=y_0$ as $y_0 \rightarrow \infty$. Note that in all these coordinates, the $(t/\tau/T,\phi/\varphi/\Phi)$ part can always be separated, due to stationarity and axi-symmetry.

The above choice essentially means that the mirror is at the Einstein static universe at the boundary of AdS. We can relax this assumption and declare that the reflecting boundary conditions are defined in terms of the Boyer-Lindquist radial coordinate instead of the Henneaux-Teitelboim radial coordinate. This will not be a conventional ``AdS/CFT" boundary condition, and the Green functions defined this way can differ from conventional AdS/CFT bulk correlators in their precise dependence on $\theta$ (see the coordinate change (\ref{HT})). Since our final goal is to gain some insight into the near-region of Kerr black holes and since the precise boundary conditions are not too crucial as long as they are reasonable and consistent, this will not deter us. So in the following, we will work with reflective boundary conditions at Boyer-Lindquist $r \rightarrow \infty$. We emphasize that at the level that we are working with, we can go back and forth between the two boundary conditions if we replace (in what follows) $\sum_l Y_l(\theta)X_{\omega_-,l,p}(r)$ with $Z_{\omega_-,p}(y, \Theta)$ where the $Z$ are (possibly inseparable in $y$ and $\Theta$) modes in the Henneaux-Teitelboim coordinates. To do this in practice though, will require us to numerically solve the partial differential equations in $y, \Theta$, with boundary conditions in $y$ rather than $r$.

With this understanding, now we present the Bogolubov transformations that take us to the Hartle-Hawking modes. For this, as in the Rindler wedge, we introduce two sets of modes on the left and right halves of the bifurcate horizon
\bea
u^{(1)}_{\omega_-, p,l} = \left\{ \begin{array}{ll}
                    \frac{1}{\sqrt{2\omega_-}}e^{-i\omega_- T+i p \Phi}Y_l(\theta)X_{\omega_-,p,l}(r)
&\quad \mbox{Right Wedge}\\
          0 & \quad \mbox{Left Wedge}
                \end{array}\right. \\
u^{(2)}_{\omega_-, p,l} = \left\{ \begin{array}{ll}
                    0
&\quad \mbox{Right Wedge}\\
          \frac{1}{\sqrt{2\omega_-}}e^{-i\omega_- T+i p \Phi}Y_l(\theta)X_{\omega_-,p,l}(r) & \quad \mbox{Left Wedge}
                \end{array}\right.
\eea
Defining
\bea
\theta_H(\omega_-)=\frac{1}{2}\log \left({e^{\beta_H \omega_-}+1\over e^{\beta_H \omega_-}-1 }\right)
\eea
the standard form \cite{Birrell, Liu} of the Bogolubov transformations is
\bea
v^{(1)}_{\omega_-, p,l}=\cosh \theta_H u^{(1)}_{\omega_-, p,l}+\sinh \theta_H u^{(2)}_{\omega_-, p,l}\\
v^{(2)}_{\omega_-, p,l}=\cosh \theta_H u^{(1)*}_{\omega_-, p,l}+\sinh \theta_H u^{(2)*}_{\omega_-, p,l}
\eea
The Hartle-Hawking modes are defined by the expansion
\bea
X(T,\Phi,\theta;r)=\sum_{\omega_-,l,p,i}(b^{(i)}_{l,\omega_-,p}v^{(i)}_{l,\omega_-,p}+b^{(i)\dagger}_{l,\omega_-,p}v^{(i)*}_{l,\omega_-,p}),
\eea
where $i$ runs over $(1,2)$ and $\omega_-$ is restricted to run only over positive values. The sum is in practice an integral. The creation-annihilation operators $b, b^\dagger$ satisfy the standard commutation relations, we follow the conventions of Birrell and Davies \cite{Birrell}. The Hartle-Hawking two point function is defined as
\bea
G_+(T-T',\Phi-\Phi',\theta,\theta';r,r')\equiv \langle0|X(T,\Phi,\theta;r)X(T',\Phi',\theta';r)|0\rangle_H \hspace{0.2in} \\
=\sum_{\omega_-,l,p} {1 \over 2 \omega_-} \ e^{ip(\Phi-\Phi')}Y_l(\theta)Y_l^*(\theta')X_{\omega_-,p,l}(r)X_{\omega_-,p,l}(r') \left[{e^{-i \omega_- (T-T')} \over 1- e^{-\beta_H \omega_-}} +{e^{i \omega_- (T-T')} \over e^{\beta_H \omega_-}-1}\right], \label{HHe}
\eea
were the vacuum $| 0 \rangle_H$ is the Hartle-Hawking vacuum defined by the condition that it is annihilated by the $b$'s. The expression on the second line here is somewhat schematic: the precise normalizations and the choice of integrations vs. sums have to be made according to the precise nature of the harmonics and their orthonormality properties. We have assumed above that the radial modes are real. Using he fact that the $\omega_-$ are positive definite, now we can go to the momentum space correlators:
\bea
G_+(\omega_-,l,p;r,r')=\frac{1}{2\omega_-}{e^{\beta_H\omega_-}\over e^{\beta_H\omega_-}-1 }X_{\omega_-,p,l}(r)X_{\omega_-,p,l}(r'). \label{momHH}
\eea
In doing these computations, we assume that the two radii, $r$ and $r'$ are on the same half-wedge. The corresponding Green function when they are on opposite wedges is computed entirely analogously \cite{Liu}:
\bea
G_{12}(\omega_-,l,p;r,r')=e^{-{\beta_H \omega_- \over 2}}G_+(\omega_-,l,p;r,r'). \label{b1b2}
\eea

When these expressions are re-interpreted in terms of the original (non co-rotating) Boyer-Lindquist or Henneaux-Teitelboim coordinates, we find that they are in the grand canonical ensemble. This gives a legitimate realization of the original argument in \cite{HH1}. To see this,
note that $G_+(\Delta T, \Delta \Phi) =  G_+(\Delta T-i\beta_H, \Delta \Phi)$, from the explicit form (\ref{HHe}) above (We have suppressed the remaining coordinates for convenience). When rewritten in terms of BL or HT coordinates, this periodicity translates to $G_+(\Delta t, \Delta \phi) =  G_+(\Delta t-i\beta_H, \Delta \phi-i\beta_H \Omega)$ and $G_+(\Delta \tau, \Delta \varphi) =  G_+(\Delta \tau-i\beta_H, \Delta \varphi-i\beta_H \Omega_0)$, which are the appropriate KMS periodicities expected in a grand canonical ensemble at temperature $1/\beta_H$ and chemical potential for the angular momentum (namely the angular velocity) $\Omega$ and $\Omega_0$ respectively. In the asymptotically non-rotating AdS frame of Henneaux and Teitelboim, we find therefore that the black hole is a thermal state described by $1/\beta_H$ and $\Omega_0$. Thermal Wightman functions defined as above have a natural entangled interpretation as well \cite{Israel, Maldacena}. The way to realize this is to define the thermal state by
\bea
|0\rangle_{\beta_H\ \Omega_0}=\frac{1}{Z^{1 \over 2}}\sum_n e^{-\beta_H(E_n-\Omega_0 J_n)/2}|E_n,J_n\rangle \otimes |E_n,J_n\rangle, \ \ {\rm where} \ Z={\rm Tr}(e^{-\beta_H(H-\Omega_0 J)}),
\eea
and then expectation values for operators on one of the Hilbert spaces can be easily seen to be thermal  expectation values in the grand canonical ensemble. This gives a natural interpretation for the two halves of the bifurcate Killing horizon \cite{Maldacena}.

\section{$AdS_3$}

The construction in the previous section was presented in the context of $AdS_{d+1}$ with $d=3$. But it is evident that it has obvious generalizations to other dimensions. Unfortunately, while the construction is conceptually well-defined in contrast to flat space, many of the technical difficulties of flat space Kerr black holes are still present in all dimensions $d \ge 3$. To make matters worse, it is not clear whether the wave equation is separable in the asymptotically AdS coordinate system of Henneaux and Teitelboim. So in this section we turn instead to the case when $d=2$, where the black hole reduces to the celebrated BTZ case \cite{BTZ}. BTZ has the wonderful advantage that the (radial) wave equation is exactly solvable, so this enables us to write down explicit Green functions using the approach of the last section. We start with some general comments.
\begin{itemize}
\item There are two major advantages to BTZ when compared to higher dimensions. One is that the radial wave equation is exactly solvable \cite{2, 3, CK1} in terms of hypergeometric functions. Secondly, since the black hole is three dimensional the only coordinates are $(t,r,\phi)$ and the spatial section of the horizon is a circle. This means that there are no extra angles one has to worry about when separating the Klein-Gordon equation, along with which comes the advantage that the precise boundary conditions on the mirror at infinity are simple.
\item Related to the previous comment is the fact that the standard BTZ coordinates for the metric (see, eg., appendix of \cite{CK1}) comes automatically in the (analogue of the) Henneaux-Teitelboim coordinate system. Since the wave equation separates already in these coordinates, one does not have to go to a Boyer-Lindquist form at all\footnote{See appendix of \cite{CK1} for the complete coordinate change that takes BTZ to a Boyer-Lindquist form.}. This enables us to work with asymptotically AdS boundary conditions while bypassing the polar angle complications in the shape of the mirror. The absence of the extra angle also means that the wave equation is separable for BTZ in either of these coordinates.
\item BTZ is a quotient of $AdS_3$. This means one can construct BTZ Green functions by starting with the covering space and using the method of images. This is the standard way in which BTZ Green functions have historically been constructed \cite{Steif,1,2,3}. It is also known that the resultant Green function is analytic in the upper half plane of the past horizon ${\cal H}^-$ and in the lower half plane of the future horizon ${\cal H}^+$, suggesting that they are Hartle-Hawking Green functions \cite{2}. This suggests that we should be able to reproduce the quotient construction result from our co-rotating construction (which is manifestly of the Hartle-Hawking form due to the Unruh-type Bogolubov transformation that we used \cite{Ross}). This would be straightforward, except  the quotient construction is in spacetime while our construction gives the results more naturally in momentum space. Doing the Fourier transform directly on the bulk Green functions seems complicated, so what we will do is to take the correlators to the boundary and then do the transform there and see whether the results match. The structure of the correlator is sufficiently involved that even there, we need some old-school integral formulae to get our answer. Fortunately, at the end of the day the results match as expected.
\item Note that there are no Kerr black holes in flat space vacuum gravity in 2+1 dimensions. BTZ black hole is possible only in AdS.  It should also be noted that since BTZ is a quotient of $AdS_3$, the local curvature is always a constant, and the singularity is an orbifold instead of a curvature singularity. For a full understanding of non-trivial local curvature effects, one has no choice but to confront the higher dimensional Kerr-AdS problem.
\end{itemize}

It is worth emphasizing that different boundary conditions give rise to different Green functions which can take entirely different functional forms. This affects us at two levels. Firstly, there is the question of what boundary conditions should we put on the covering AdS space in constructing the Green functions. Secondly, once such a choice is made in the covering space and the quotient Green functions are constructed, one needs to clarify what are the boundary conditions that are satisfied by these Green functions in the quotient space (i.e, on BTZ).

In particular, the first paper to do the covering space approach was by Steif 
\cite{Steif}. But the boundary conditions adopted there correspond to what are called transparent boundary conditions in the language of \cite{Avis}. These are not the most natural choice from an AdS/CFT point of view. Indeed, later the same images approach was applied by \cite{1, 2} with the conditions that there is no energy leakage at the boundary of the covering space, i.e., reflecting  boundary conditions. The quotient Green functions constructed this way are what are usually called BTZ Green functions \cite{3, Levi, Bala} in the AdS/CFT context. These are what we will use and it is this choice that leads naturally to the Hartle-Hawking boundary conditions in the quotient space. This demonstrates that this choice is a canonical one, both from the covering space point of view as well as the black hole point of view.

Now we adapt the general results of the previous section to the specific case of BTZ and write down explicit correlators. We will follow the notations of \cite{CK1}. The standard BTZ form of the metric is
\bea
ds^2=-\frac{(r^2-r_+^2)(r^2-r_-^2)}{r^2} dt^2+\frac{r^2}{(r^2-r_+^2)(r^2-r_-^2)}dr^2+r^2
\Big(d\phi-\frac{r_-r_+}{r^2}dt\Big)^2. \label{BTZ}
\eea
One can go to the co-rotating frame by \cite{Levi, Bala, CK1}
\bea
T=t, \ \ \Phi=\phi-\Omega_0 t, \ \ {\rm where} \ \ \Omega_0={r_- \over r_+}.
\eea
As advertised, the standard BTZ coordinate system is analogous to the Henneaux-Teitelboim coordinates and therefore is asymptotically locally AdS and static. We seek scalar mode solutions in the form
\bea
u_{\omega_-,p}(T,\Phi;r)=
\frac{1}{\sqrt{2\omega_-}}e^{-i\omega_- T+i p \Phi}X_{\omega_-,p}(r).
\eea
The wave equation for a massive scalar of mass $m$ in the co-rotating frame takes the form
\bea
X_{\omega_-, p}''+\frac{(rN^2)'}{rN^2}X_{n\omega_-}'+ \nonumber \hspace{4.3in}\\
\hspace{0.5in}+\frac{1}{r^2N^4}
\Big[r^2\Big(\big(\omega_-+{r_-\over r_+}p\big)^2-p^2\Big)-2r_-r_+\omega_- p -p^2{(r_+^2-r_-^2)}-m^2 r^2 N^2\Big]X_{\omega_-, p}=0,
\eea
with the lapse defined as
\bea
N^2(r)={(r^2-r_+^2)(r^2-r_-^2) \over r^2}.
\eea
By a simple adaptation of previous work\footnote{See  the computations in section 4.2 of \cite{CK1}. The results there can be brought to the set-up we are working with by simple variable changes.}, we can solve this with normalizable AdS fall-offs at the boundary. The result is
\bea
X_{\omega_-,p}(r)&=&C(\omega_-,p)(u-1)^\alpha
u^{-\alpha-h_+}F(\alpha+\beta+h_+,\alpha-\beta+h_+;1+\nu;1/u).
\eea
with the new radial variable $u$ defined as
\bea
u={r^2-r_-^2 \over r^2-r_+^2}.
\eea
The coefficient $C(\omega_-,p)$ given by
\bea
C(\omega_-, p)^2=
\frac{\Gamma(h_+-\alpha-\beta)\Gamma(h_+-\alpha+\beta)\Gamma(h_++\alpha+
\beta)\Gamma(h_++\alpha-\beta)}{\Gamma(1+\nu)^2\Gamma(2\alpha)\Gamma(-2\alpha)}
\eea
is an important quantity in the following discussion.  The other symbols are defined as follows. The function $F$ stands for the hypergeometric function ${}_2F_1$, while
\bea
\nu =\sqrt{1+m^2}, \ \ h_{+}={1+ \nu \over 2}, \ \ \alpha=i{ r_+ \omega_- \over 2(r_+^2-r_-^2)}, \ \ \beta=i{\Big( r_- \omega_--{(r_+^2-r_-^2)\over r_+}p \Big)\over 2(r_+^2-r_-^2)}
\eea
With this solution, we finally have the explicit bulk Hartle-Hawking Green function by directly substituting the above expressions into (\ref{momHH}) (one should suppress the $l$ index because BTZ has no polar angle). The result is a complicated expression which is trivial to write down using our results, but we will not write it down to avoid clutter. We can also define boundary correlators by the usual AdS/CFT correspondence \cite{adscft} from these bulk Green functions. We will follow the conventions of \cite{Liu, CK1} in going to the boundary\footnote{There is an extra factor of $4 \nu^2$ in the boundary correlators in the notations of \cite{Liu, CK1} (see Eqn. (3.3) in \cite{CK1}) when compared to some other conventions in the literature. But this factor appears both in the construction here as well as the one based on the quotient space, so does not affect the comparison.}. Then, because the hypergeometric functions simplify, the boundary1-to-boundary2 correlators take the simpler, but still complicated, form  (see (\ref{b1b2})):
\bea
G_{12}(\omega_-,p)=\frac{\beta_H (r_+^2-r_-^2)^{1+\nu} 
}{2\pi^2 \Gamma(\nu)^2} \Gamma(h_+-\alpha-\beta)\Gamma(h_+-\alpha+\beta)\Gamma(h_++\alpha+
\beta)\Gamma(h_++\alpha-\beta). \nonumber \\ \label{b1b2n}
\eea
Here $\beta_H$ is the inverse Hawking temperature of the BTZ black hole given by
\bea
\beta_H= {2 \pi r_+ \over r_+^2-r_-^2}.
\eea
The structure of the Gamma functions here has a striking similarity with the Kerr-CFT absorption cross-sections written down by CMS \cite{CMS} (see also \cite{CK2, rest} which generalized the idea to more general black holes.) The expression for the $C(\omega_-,p)^2$ above and the expression (6.11) in CMS are analogous when one recalls Euler's reflection formula $\Gamma(ix)\Gamma(-ix)\sim {1 \over \sinh \pi x}$. As we will soon show, the BTZ Green functions here can  be obtained by starting from the covering $AdS_3$. In the Kerr-CFT case also, there is an auxiliary $AdS_3$ that shows up in the near-region wave equation, which is broken by the periodic identification in $\phi$. 

It is clear that the analogy is far from perfect. Firstly, it should be kept in mind that here we are computing Green functions, whereas CMS considered cross-sections (hence the absolute values in the expressions). Some related discussion can be found in sections 2.12 and 3.1 of \cite{MaldStrom}. Another comment is that the temperature arising for BTZ is the Hawking temperature, while there are two notions of temperature (left and right) in the case of the Kerr black hole and they are not (at least {\em a priori}) related to the Hawking temperature. 

Despite these difficulties, it seems evident that the absoprtion cross sections for Kerr-CFT should be thought of as descending from the auxiliary $AdS_3$ in parallel to the fact that Green functions on BTZ arose from  the covering $AdS_3$. An approach to generate the scattering amplitudes in \cite{CMS} from the near-region $AdS_3$ is currently under way. 

Now we turn to the promised demonstration that the Green functions above can indeed be obtained from the  covering $AdS_3$ space. The construction with the appropriate no leakage boundary conditions in the covering space can be found in \cite{1, 2}. In \cite{2}, the Feynman Green functions were constructed, starting with the covering space result given by their Eqn. (A.22). What we need are Wightman Green functions. Fortunately, this is easy because the two are related by
\bea
G_F(t)=\theta(t)G_+(t)+\theta(-t)G_-(t)
\eea
and it is straightforward to adapt their work. Using $f(x)\theta(x)+f(-x)\theta(-x)=f(|x|)$ this essentially just removes an absolute value sign in their expressions. Using this and repeating their method of images construction we arrive at the bulk Green function on the quotient (BTZ) space, which is essentially identical to (4.5-4.7) in \cite{2}:
\bea
G_{\rm BTZ}^{\rm bulk}(\Delta T, \Delta \Phi; r,r')\sim \frac{1}{4\pi}\sum_{n=-\infty}^{\infty}{[z_n+(z_n^2-1)^{1/2}]^{1-2 h_+}\over (z_n^2-1)^{1/2}}. \label{q1}
\eea
We put $\sim$ sign rather than an equality because we do not want to keep track of some constant normalizations.  The $z_n$ are defined in terms of the BTZ variables as
\bea
z_n=\frac{1}{r_+^2-r_-^2}\Big(\sqrt{(r^2-r_-^2)(r'^2-r_-^2)}\cosh(r_+ \Delta \Phi_n)+\hspace{1.5in}\nonumber \\ \hspace{1.5in}+\sqrt{(r^2- r_+^2)(r'^2-r_+^2)}\cosh({(r_+^2-r_-^2)\over r_+} \Delta T-r_-\Delta \Phi_n)\Big), \label{q2}
\eea
with $\Delta \Phi_n \equiv \Delta \Phi+2 \pi n$.
The $r$ and $r'$ are on different asymptotic regions (our aim is to reproduce (\ref{b1b2n})). This results in a shift by $-i\beta_H/2$ in $\Delta T$ compared to the case when $r$ and $r'$ are on the same region, see appendix B of \cite{CK1}. In any event, (\ref{q1})-(\ref{q2}) together provide us the expression for the bulk Green function in the quotient approach. When we take it to the boundary, this reduces to a form that is a version of the well-known \cite{3, Levi, Bala} result
\bea
G_{12}(\Delta T, \Delta \Phi)\sim \frac{4 \nu^2 (r_+^2-r_-^2)^{1+\nu}}{2 \pi}\sum_n \left(\frac{1}{e^{r_+\Delta  \Phi_n}+e^{-r_+\Delta \Phi_n}+  e^{a_H \Delta T-r_-\Delta \Phi_n}+e^{-(a_H \Delta T-r_-\Delta \Phi_n)} }\right). \nonumber \\ \label{b1b2nn}
\eea
We use the shorthand $a_H=2 \pi/ \beta_H$. The boundary correlator has the advantage of being manageable.

We claim that
\bea
G_{12}(\omega_-,p)=\int_{-\infty}^{\infty}d\Delta T \int_{0}^{2 \pi}d \Delta \Phi \ e^{-i \omega_- \Delta T+i p \Delta \Phi} G_{12}(\Delta T, \Delta \Phi).
\eea
To check this, we need an integral that is provided in the appendix, together with some minor further change of variables. The end result is that, modulo the constant normalizations which we have not kept track of, the two results (\ref{b1b2n}) and (\ref{b1b2nn}) match precisely. This is a match involving functional forms and various parameters and is therefore quite non-trivial.

The quotient construction provides legitimacy to the co-rotating construction of the Hartle-Hawking Green function on BTZ. The quotient Green function is a canonical one from the perspective of $AdS_3/CFT_2$ and string theory, and has been put to considerable study. In higher dimensions where the quotient approach is unavailable, the co-rotating approach emerges as the natural candidate for constructing thermal Hartle-Hawking-like Green functions.

\section*{Acknowledgments}

I would like to thank Daniel Arean, Cyril Closset, Jarah Evslin,  Josef Lindman-Hornlund, Carlo Maccaferri, Bindusar Sahoo and  Shahin Sheikh-Jabbari for interesting discussions and Guido Festuccia, Bernard Kay and Per Kraus for helpful correspondence.

\section{Appendix}
\subsection*{{\bf A.} \ An Integral}
\addcontentsline{toc}{subsection}{{\bf A} \ An Integral}
\renewcommand{\theequation}{A.\arabic{equation}}

\begin{figure}
\begin{center}
\includegraphics[height=0.4\textheight
]{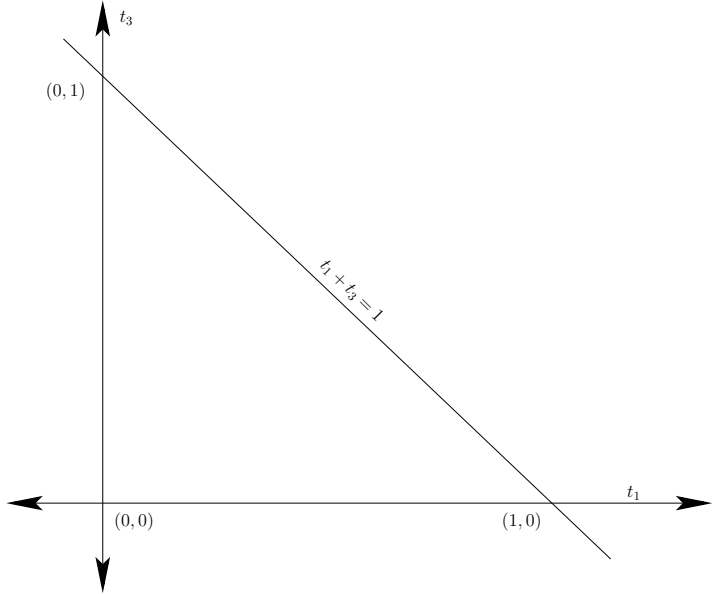}
\caption{The integration is in the region inside the triangle.}
\label{range}
\end{center}
\end{figure}

The following integral is useful in the main text:
\bea
I(\omega,p)=\int_{-\infty}^{\infty}\int_{-\infty}^{\infty}\frac{e^{-i\omega t+i p \phi}}{(e^{\phi/2}+e^{-\phi/2}+e^{t/2}+e^{-t/2})^{\lambda}}dt \ d\phi
\eea
We have not been able to do this integral using Mathematica, so we will use the following substitution:
\bea
t_1=\frac{e^{\phi/2}}{S}, \ t_2=\frac{e^{-\phi/2}}{S}, \ t_3=\frac{e^{t/2}}{S}, \ t_4=\frac{e^{-t/2}}{S}
\eea
with $S=(e^{\phi/2}+e^{-\phi/2}+e^{t/2}+e^{-t/2})$.  Not all $t_i$ are independent:
\bea
t_1 t_2=t_3 t_4, \ \ t_1+t_2+t_3+t_4=1.
\eea
We choose to solve for $t_4$ and $t_2$ in terms of $t_1, t_3$:
\bea
t_2=\frac{t_3(1-t_1-t_3)}{(t_1+t_3)}, \ \ t_4=\frac{t_1(1-t_1-t_3)}{(t_1+t_3)}, \  \ S=\left(\frac{t_1+t_3}{t_1t_3(1-t_1-t_3)}\right)^{1/2}.
\eea
To express the integral in terms of the new independent variables ($t_1, t_3$) we need to understand the ranges that $t_1$ and $t_3$ sweep out as $t$ and $\phi$ range from $-\infty$ to $+\infty$, and also the Jacobian for the transformation. To fix the ranges, it is easier to work with
\bea
e^{(\phi+t)/2}=\frac{t_1+t_3}{1-t_1-t_3}, \ \ e^{(\phi-t)/2}=\frac{t_1}{t_3},
\eea
both of which should range from $0$ to $\infty$. It is easily seen that this is achieved by integrating $t_1$ and $t_3$ between $0$ and $1$ such that $t_1+t_3$ runs between $0$ and $1$. The range of the integration is given in Figure \ref{range}.
The Jacobian is easily computed:
\bea
|J|=\frac{2}{t_1t_3(1-t_1-t_3)}
\eea
The end result is that the integral $I(\omega, p)$ now takes the form
\bea
I(\omega, p)= 2 \int_R dt_1 dt_3  \frac{ t_1^a t_3^b(t_1+t_3)^c}{(1-t_1-t_3)^{1+c}}
\eea
where $R$ is the range given in the figure and
\bea
a=ip+i\omega+\frac{\lambda}{2}-1, \ \ b= -i p-i\omega+\frac{\lambda}{2}-1, \ \ c= ip -i\omega-\frac{\lambda}{2}.
\eea
Fortunately, this integral is of a standard form (See formula 4.635 (2) in  \cite{Gradshteyn}). Using also the result that $\int_0^1 dx \  x^m/(1-x)^n=\Gamma(1+m)\Gamma(1-n)/\Gamma(2+m-n)$ we finally end up with
\bea
I(\omega,p)=\frac{2}{\Gamma(\lambda)^2}\Gamma\Big(ip+i\omega+\frac{\lambda}{2}\Big)\Gamma\Big(-i p-i\omega+\frac{\lambda}{2}\Big)\Gamma\Big(-i p+i\omega+\frac{\lambda}{2}\Big)\Gamma\Big(i p-i\omega+\frac{\lambda}{2}\Big)
\eea
%

%

\end{document}